\documentclass[aps,pre,superscriptaddress,preprint,showkeys,showpacs,floatfix]{revtex4}
\usepackage{epsfig}
\usepackage{amsmath,amssymb,amsfonts,latexsym}
\usepackage{graphicx}

\usepackage{epsfig,amssymb,amsfonts,verbatim}
\usepackage{amsmath}
\usepackage{latexsym}
\usepackage{amsfonts}
\usepackage{amssymb}
\usepackage{color}
\def\bfl{\begin{flushleft}}
\def\efl{\end{flushleft}}
\def\bfr{\begin{flushright}}
\def\efr{\end{flushright}}
\def\bc{\begin{center}}
\def\ec{\end{center}}

\def\ba{\begin{eqnarray}}
\def\ea{\end{eqnarray}}
\def\baa#1{\begin{array}{#1}}
\def\eaa{\end{array}}
\def\bw{\begin{widetext}}
\def\ew{\end{widetext}}

\def\text#1{\mbox{#1}}

\Large
\begin{document}

\title{Supercurrent flow with large superconductor gap in cuprates: Resurrection of phonon-mediated Cooper pairs}
\author{Andrew Das Arulsamy}
\email{sadwerdna@gmail.com}
\affiliation{Condensed Matter Group, Institute of Interdisciplinary Science, No.~24, level-4, Block C, Lorong Bahagia, Pandamaran, 42000 Port Klang, Selangor DE, Malaysia}

\date{\today}

\begin{abstract}
We systematically explore the exquisiteness of Bardeen-Cooper-Schrieffer(BCS) Hamiltonian where the BCS-type electron-phonon interaction is unambiguously reinforced as the only viable superglue in cuprate superconductors because phonon-induced scattering is effectively nil for Cooper pairs (in its original form), and also phonons are never required to Bose-condense. Here, we prove that (i) the Cooper-pair binding energy can be strengthened to obtain high superconductor transition temperature ($T_{\rm sc}$) and (ii) the existence of a generalized electron-phonon potential operator that can induce the finite-temperature quantum phase transition between superconducting and strange metallic phases. To lend support for this extended BCS Hamiltonian, we derive the Fermi-Dirac statistics for Cooper-pair electrons, which correctly captures the physics of strongly bounded Cooper-pair break up with respect to changing temperature or superconductor gap ($\Delta^{\rm BCS}$). Finally, we further extend the BCS Hamiltonian within the ionization energy theory formalism to prove (iii) the existence of optimal doping that has maximum $T_{\rm sc}(x_{\rm optimum})$ or $\Delta^{\rm BCS}(x_{\rm optimum})$, and (iv) that the specific heat capacity jump at $T_{\rm sc}$ in cuprates is due to finite-temperature quantum phase transition. Along the way, we expose the precise microscopic reason why predicting (not guessing) a superconductor properly is a hard problem within any theory that require pairing mechanism.  
\end{abstract}



\maketitle

\section*{1. Introduction}

Above zero Kelvin, phonons do not Bose-condense, they never did, and consequently, ions are found to vibrate independently in every corner of a crystal. Hence, phonons define the ultimate resistance to electrons or bosons flow, in which, this resistance does not go away until we reach absolute zero. It is a fundamental fact that for any conducting particles or pairs of particles (namely, Cooper pairs, bipolarons, holon pairs, anyons or bosons) to superconduct, they eventually need the `approval' of phonons such that the particle-phonon or paired particle-phonon scattering is completely removed. The only theory that properly and correctly eliminates the phonon-induced scattering is the BCS (Bardeen, Cooper and Schrieffer) Hamiltonian such that phonons are never required to Bose condense~\cite{bard}. 

Strangely, BCS Hamiltonian of superconductivity on the basis of Cooper pairs~\cite{coop} has been abruptly sidelined after the discovery of cuprate superconductors (also known as the high temperature superconductors) by Bednorz and M${\rm \ddot{u}}$ller (BM)~\cite{norz}. The argument is that BCS Hamiltonian is only applicable for weakly coupled Onnes-type~\cite{onn} conventional superconductors with $s$-wave pairing. Here, the notion of weak coupling refers to the strength of electron-phonon (e:ph) coupling that is responsible for the formation of Cooper pairs, which is found to be too weak (with large coherence length) to produce high superconductor transition temperature ($T_{\rm sc}$). The highest BCS transition temperature, $T^{\rm BCS}_{\rm sc}$ is about 40 K in MgB$_2$ superconductor~\cite{mgb2}, while Hg-based cuprate has the highest BM transition temperature ($T^{\rm BM}_{\rm sc}$), which is about 130 K~\cite{bm}. Hence, other types of pairing mechanisms have been proposed by ignoring the phonon-induced scattering effect. 

On the basis of general consensus (somewhat similar to the enforced Copenhagen interpretation), two alternative proposals have been `elected' because they are supported by certain experiments, and only one of them is believed to hold the key ingredients for high $T^{\rm BM}_{\rm sc}$ superconductivity~\cite{mann}. Briefly, the elected proposals are---(i) the magnetic spin fluctuation induced superglue~\cite{pines} and (ii) the Anderson resonating valence bond theory that require Bose-Einstein condensation of holon pairs and spin-liquid~\cite{ander,basko,bask,ede}. For some in-depth arguments in favor of these proposals, refer to the reviews written by Baskaran~\cite{bask} and Scalapino~\cite{scala}.

Here, we do not follow any of these alternatives for two theoretically solid reasons---the first has been exposed earlier (see the first paragraph), while the second reason shall be explained in the following paragraphs. Note this, BCS Hamiltonian is not dead and buried for unconventional superconductors because the correctness and validity of any theory should never be based on democracy. Instead, all microscopic theoretical mechanisms should be properly verified with the most relevant experiments and low-level analytic analysis (to show that there is no internal inconsistency). Here, `low-level' means at the `operator level' where the operators themselves are subjected to formal analytic and theoretical abuse to check for their internal consistency. 

Apart from that, unlike resistivity~\cite{onn}, Meissner-effect~\cite{meiss} and specific heat capacity ($C_{\rm v}$) measurements~\cite{corak}, the analysis based on ARPES (Angle-Resolved Photoemission Spectroscopy), tunneling and nuclear magnetic resonance (NMR) measurements~\cite{timu,manne} cannot be used (on their own) to unambiguously deduce the existence of superconductivity, unless one is already aware that the system is a superconductor. On the other hand, the meaning of this statement---`microscopic mechanism without internal inconsistency' shall be exposed when we revisit the groovy BCS Hamiltonian in its full glory (see Eqs.~(\ref{eq:1}),~(\ref{eq:2}) and~(\ref{eq:3})). 

The Cooper-pair mechanism (in its original form) has been abandoned, or presently, one is forced to do so enforced by the general consensus on the basis of the following weak arguments. The binding energy of Cooper pairs (popularly known as the superconductor gap) cannot be made large enough to transform $T^{\rm BCS}_{\rm sc} \rightarrow T^{\rm BM}_{\rm sc}$ because the BCS mechanism of phonon-mediated Cooper pairing is limited to $s$-wave pairing and also due to small $|E_{\textbf{k}} - E_{\textbf{k} + \textbf{q}}|$ where $E_{\textbf{k}} > \hbar\omega_{\textbf{q}}$, $E_{\textbf{k} + \textbf{q}} > \hbar\omega_{\textbf{q}}$ and $|E_{\textbf{k}} - E_{\textbf{k} + \textbf{q}}| \ll \hbar\omega_{\textbf{q}}$. The first two inequalities allow adiabatic approximation due to this time relation, $t_{\rm e} < t_{\rm ph}$ (an electron responds at a faster timescale compared to a phonon decay), which is crucial for the formation of Cooper pairs. Here, $|E_{\textbf{k}} - E_{\textbf{k} + \textbf{q}}| \cong k_{\rm B}T^{\rm BCS}_{\rm sc}$ where $k_{\rm B}T^{\rm BCS}_{\rm sc} = \Delta^{\rm BCS}$ denotes the superconductor gap. The second inequality defines the attraction (if $|E_{\textbf{k}} - E_{\textbf{k} + \textbf{q}}| < \hbar\omega_{\textbf{q}}$) between two electrons such that one of the electrons has a change of energy from $E_{\textbf{k}}$ to $E_{\textbf{k} + \textbf{q}}$, while the energy of the second electron changes from $E_{\textbf{k}'}$ to $E_{\textbf{k}' - \textbf{q}}$ mediated by a phonon absorption and emission, respectively, with energy $\hbar\omega_{\textbf{q}}$. 

BCS happened to derive the gap equation and the $C_{\rm v}$ relation based on $s$-wave pairing (spherical Fermi surface). But this is never a restriction for the application of BCS Hamiltonian in cuprates because their Hamiltonian permits singlet pairing for whatever Fermi surfaces, $s$- or $p$- or $d$- or $f$-wave pairing symmetry, or any combination of them. It is just a matter of finding which atoms in a given superconductor contribute to Cooper pairing. For example, light atoms may give rise to $s$- or $p$-wave singlet pairing, whereas heavier atoms may lead to $d$- or $f$-wave singlet pairing. Additionally, the types of atoms and their sequence found along the $a$, $b$ or $c$ axis in cuprates are different. This means that, the existence of anisotropic (or quasi-two dimensional) normal state resistivity ($ab$-plane versus $c$-axis) above $T_{\rm sc}$ is expected because the normal state is not a free-electron or Fermi-liquid metal. But this quasi-two dimensional conductivity does not invalidate the formation of BCS-type Cooper pairs. When one comes to think of the reasons stacked against the BCS Hamiltonian, one has no other option but to sound heretical for the stacked reasons are scientifically lame. Here, the term Cooper-pair strictly refers to the original BCS-type, and we do not refer to any other types. More details on the $d$-wave pairing symmetry are available in Ref.~\cite{scalap}. 

The existence of pseudogap for $T^* > T_{\rm sc}$ and doping-dependent superconductor gap ($\Delta^{\rm BCS}$) can be made to obey the conduction mechanism of Cooper pairs by reworking the BCS attraction operator within the ionization energy theory (IET)~\cite{pra}. In fact, IET has enabled us to address the strange metallic phase and doping-dependent resistivity~\cite{jsnm,jsnm2} above $T_{\rm sc}$ properly. However, we have to postpone the research on pseudogap because BCS Hamiltonian, even after extension, does not `uniquely' lead us to find the origin of this gap. For example, we can always assume that preformed Cooper pairs or some forms of phonon `readjustment' is the cause for this pseudogap where both can be related to BCS Hamiltonian. Warning: Cooper-pair formation is responsible for an upward (not downward) `jump' in the $C_{\rm v}$ data at $T_{\rm sc}$, and therefore, preformed Cooper pairs (if they really exist) should be detectable with a similar upward-jump (however small) in the $C_{\rm v}$ measurements for $T > T_{\rm sc}$. Thus far, there is no such data reported.

We now introduce the rituals needed to resurrect Cooper pairs within BCS Hamiltonian, including the reasons for the resurrection. First, let us recall the BCS Hamiltonian that takes the phonon assisted electron-electron (e:e) attraction into account (for singlet pairing), which is given by~\cite{bard}
\begin {eqnarray}
&&H_{\rm BCS} = \sum_{k > k_{\rm F}}E_{\textbf{k}}n_{\textbf{k}\sigma} + \sum_{k < k_{\rm F}}|E_{\textbf{k}}|(1 - n_{\textbf{k}\sigma}) + H^{\rm screened}_{\rm Coulomb} + H^{\rm Cooper}_{\rm pair}, \label{eq:1} \\&& H^{\rm screened}_{\rm Coulomb} = \sum_{\textbf{k}}\frac{e^2}{\epsilon_0(\textbf{k}^2 + K_s^2)}\label{eq:2}, \\&& H^{\rm Cooper}_{\rm pair} = \frac{1}{2}\sum_{\textbf{k},\textbf{k}',\sigma,\sigma',\textbf{q}}\frac{2\hbar\omega_\textbf{q}|g_{\textbf{k},\textbf{k}'}|^2c^*(\textbf{k}'-\textbf{q},\sigma')c(\textbf{k}'\sigma')c^*(\textbf{k}+\textbf{q},\sigma)c(\textbf{k},\sigma)}{(E_{\textbf{k}} - E_{\textbf{k} + \textbf{q}})^2 - (\hbar\omega_{\textbf{q}})^2} \label{eq:3}, 
\end {eqnarray}  
where $|g_{\textbf{k},\textbf{k}'}| = |\langle\psi_{\textbf{k}'}|H_{\rm e:ph}|\psi_{\textbf{k}}\rangle|$ and $H_{\rm e:ph}$ denotes the usual e:ph interaction Hamiltonian. Here, the crystal momentum is conserved ($\textbf{k} + \textbf{k}' = (\textbf{k} + \textbf{q}) + (\textbf{k}' - \textbf{q})$) as required, the first two terms in Eq.~(\ref{eq:1}) refer to kinetic energies below and above Fermi surface ($k_{\rm F}$), $n_{\textbf{k}\sigma}$ is the electron number operator with spin, $\sigma$, $K_s$ denotes the Thomas-Fermi screening length, $\textbf{k}$ and $|\textbf{k}| = k$ are the respective wave vector and wavenumber for an electron, while $\textbf{q}$ and $\omega_{\textbf{q}}$ are the phonon wave vector and frequency, respectively. The screened Coulomb-Hamiltonian (see Eq.~(\ref{eq:2})) gives the e:e repulsion after factoring in the screening effect, $\hbar$ is the Planck constant divided by 2$\pi$, $\epsilon_0$ denotes the permittivity of free space and $e$ is the electron charge. The factor 1/2 avoids counting the same Cooper pair twice, if $\sigma$ = $\uparrow$ then $\sigma'$ = $\downarrow$, $\textbf{k}'$ = $-\textbf{k}$ such that $\textbf{k} + \textbf{k}' = \textbf{q} = 0$. The last two requirements on $\sigma' = -\sigma$ and $\textbf{q} = 0$ are to maximize $|H^{\rm Cooper}_{\rm pair}|$ defined in Eq.~(\ref{eq:3}), which are as they should be if one were to determine the superconducting ground state with efficient phonon exchange between two electrons forming a Cooper pair. 

Finally, the e:ph coupling constant or its matrix element is denoted by $g_{\textbf{k},\textbf{k}'}$, $c^*(\cdots)$ and $c(\cdots)$ are the usual electron creation and annihilation operators, respectively. For example, $c^*(\textbf{k}+\textbf{q},\sigma)$ creates an electron by absorbing a phonon after annihilating the electron (prior to absorption) with $c(\textbf{k},\sigma)$. The formation of a Cooper pair requires another electron to emit the previously absorbed phonon such that the second electron is first annihilated, $c(\textbf{k}',\sigma')$, and then recreated by emitting the absorbed phonon, $c^*(\textbf{k}'-\textbf{q},\sigma')$. Therefore, the electron with $\textbf{k}+\textbf{q}$ and $\sigma$ is paired with the second electron with $\textbf{k}'-\textbf{q}$ and $\sigma'$, and they form a Cooper pair. 

Note this, the transition to superconducting phase is readily achieved for $H^{\rm screened}_{\rm Coulomb} < |H^{\rm Cooper}_{\rm pair}|$, while preformed Cooper pairing above $T_{\rm sc}$ is possible if $H^{\rm screened}_{\rm Coulomb} > |H^{\rm Cooper}_{\rm pair}|$ and $|H^{\rm Cooper}_{\rm pair}| \neq 0 \neq E_{\textbf{k}} - E_{\textbf{k} + \textbf{q}}$. If $E_{\textbf{k}} - E_{\textbf{k} + \textbf{q}} = 0$, then $H^{\rm Cooper}_{\rm pair} = 0$ and consequently, Cooper-pair concentration is zero. Moreover, physically $(E_{\textbf{k}} - E_{\textbf{k} + \textbf{q}})^2 \geq (\hbar\omega_{\textbf{q}})^2$ is never allowed, but in any case, $H^{\rm Cooper}_{\rm pair}$ is zero by definition for this second condition. Clearly, phonon induced scattering has been eliminated because they contribute to Cooper pairing, and therefore, Bose condensation of phonons is never required. Of course, in the absence of phonons (at absolute zero), one has maximum BCS superconductor gap as a result of maximum number of Cooper pairs and efficient Cooper-pair formation because lattice distortion for $T = 0$K originates entirely from the emission and absorption of phonons by the Cooper-pair electrons, via the temperature-independent e:ph interaction.         

To extend BCS Hamiltonian to cuprates, one needs to understand the origin of Eq.~(\ref{eq:3}), which can be traced back to the derivation of the e:ph interaction term from the second-order perturbation theory and Landau's approach~\cite{aop}. In particular, the e:ph potential operator~\cite{aop},
\begin {eqnarray}
V_{\textbf{k},\textbf{k}'} = \frac{2\hbar\omega_\textbf{q}|g_{\textbf{k},\textbf{k}'}|^2}{(\hbar\omega_{\textbf{q}})^2 - (E_{\textbf{k}} - E_{\textbf{k} + \textbf{q}})^2}, \label{eq:4} 
\end {eqnarray}  
after a change of notation, $-\textbf{k}^* \rightarrow \textbf{k}'$ and removal of the energy-level spacing ($\xi$) term, $\exp{[(1/2)\lambda(\xi-E^0_{\rm F})]}$ by taking $\xi = E^0_{\rm F}$ where $\lambda = (12\pi\epsilon_0/e^2)a_{\rm B}$, $a_{\rm B}$ is the Bohr radius and $E^0_{\rm F}$ denotes the Fermi level for $T = 0$K. Here $V_{\textbf{k},\textbf{k}'}$ is positive because $(\hbar\omega_{\textbf{q}})^2 > (E_{\textbf{k}} - E_{\textbf{k} + \textbf{q}})^2$, and large e:ph interaction (or large $E_{\textbf{k}} - E_{\textbf{k} + \textbf{q}}$) leads to large e:e repulsion~\cite{pra}, which is also inevitable from Eq.~(\ref{eq:4}) where $E_{\textbf{k}}$ and $E_{\textbf{k} + \textbf{q}}$ refer to the same electron (before and after phonon absorption). This electron interacts strongly with another electron if $E_{\textbf{k}} - E_{\textbf{k} + \textbf{q}}$ is large, which is implicit from Eq.~(\ref{eq:4}). This interaction is obviously always repulsive between two different electrons, which can only be converted into an effective attraction between these electrons by switching the signs for $+(\hbar\omega_{\textbf{q}})^2$ and $-(E_{\textbf{k}} - E_{\textbf{k} + \textbf{q}})^2$ in the denominator of Eq.~(\ref{eq:4}). 

This sign-switch converts the repulsive interaction into an attraction because as stated earlier, $(\hbar\omega_{\textbf{q}})^2 \leq (E_{\textbf{k}} - E_{\textbf{k} + \textbf{q}})^2$ is physically impossible. This switch uniquely leads us to Cooper pairing mechanism where one of the electrons interact attractively with another electron via absorption (electron 1) and emission (electron 2) of phonons. In particular, the said sign-switch in Eq.~(\ref{eq:4}) naturally activates the formation of Cooper pairs via the following notions---`if' $E_{\textbf{k}} - E_{\textbf{k} + \textbf{q}}$ is true for the first electron, and `if' $E_{\textbf{k}'} - E_{\textbf{k}' - \textbf{q}}$ is also allowed to be true for the second electron, and `if' these changes in energies (for both electrons) should occur faster than the phonon timescale, $t_{\rm ph} = 1/\omega_{\textbf{q}}$, then the formation of Cooper pairs is inevitable. Here, the electrons always respond faster than phonons ($t_{\rm ph} = 1/\omega_{\textbf{q}} > t_{\rm e} = 1/E_{\textbf{k}}$) because $E_{\textbf{k}} > \hbar\omega_{\textbf{q}}$, $E_{\textbf{k} + \textbf{q}} > \hbar\omega_{\textbf{q}}$, $E_{\textbf{k}'} > \hbar\omega_{\textbf{q}}$ and $E_{\textbf{k}' - \textbf{q}} > \hbar\omega_{\textbf{q}}$. Implementing the above notions (the highlighted `if's) leads one to transform Eq.~(\ref{eq:4}) into,      
\begin {eqnarray}
V^{\rm attraction}_{\textbf{k},\textbf{k}'} = \frac{2\hbar\omega_\textbf{q}|g_{\textbf{k},\textbf{k}'}|^2}{-(\hbar\omega_{\textbf{q}})^2 + (E_{\textbf{k}} - E_{\textbf{k} + \textbf{q}})^2}. \label{eq:5} 
\end {eqnarray}  
After incorporating the creation and annihilation operators for both electrons, the factor 1/2, the sum to count all the occupied states in momentum space, and of course, after tacking the spin ($\sigma$ or $\sigma'$) for each electron as required to form strongly bounded Cooper pairs, one can obtain Eq.~(\ref{eq:3}) from Eq.~(\ref{eq:5}). The arguments used to derive Eq.~(\ref{eq:3}) from Eq.~(\ref{eq:4}) were also exploited to construct the `first paragraph'. For example, phonon-induced scattering has been eliminated without enforcing phonons to Bose-condense. Note this, much stronger Cooper pairs (large $V^{\rm attraction}_{\textbf{k},\textbf{k}'}$) can be formed (from Eq.~(\ref{eq:3}) or Eq.~(\ref{eq:5})) if one could further enhance the electron-ion attraction strength such that $(\hbar\omega_{\textbf{q}})^2 \gg (E_{\textbf{k}} - E_{\textbf{k} + \textbf{q}})^2 \rightarrow (\hbar\omega_{\textbf{q}})^2 > (E_{\textbf{k}} - E_{\textbf{k} + \textbf{q}})^2$. The second inequality shall lead us to a much smaller denominator, and therefore to a large $V^{\rm attraction}_{\textbf{k},\textbf{k}'}$ or $H^{\rm Cooper}_{\rm pair}$. The first inequality is for conventional superconductors.

On the other hand, if the above highlighted `if's are not true, then Cooper pairing is not possible, and therefore, Eq.~(\ref{eq:5}) or Eq.~(\ref{eq:3}) needs to transform in order to be superseded by Coulomb repulsion between electrons, giving rise to an insulating ground state or a strange metallic phase. In this case, Eq.~(\ref{eq:5}) or Eq.~(\ref{eq:3}) reverts to Eq.~(\ref{eq:4}), which in turn implies the existence of finite-temperature quantum phase transition (QPT$^{\rm >0K}$). Here, one is naturally led to relate this phase transition to strange metallic- to superconducting-phase transition, or more precisely, the transition from Eq.~(\ref{eq:4}) (for $T > T_{\rm sc}$) to Eq.~(\ref{eq:5}) (or Eq.~(\ref{eq:3})) below $T_{\rm sc}$. We shall comeback to this point later.

Now assuming Eq.~(\ref{eq:3}) or Eq.~(\ref{eq:5}) activates the formation of weakly coupled BCS Cooper pairs (by assuming $(\hbar\omega_{\textbf{q}})^2 \gg (E_{\textbf{k}} - E_{\textbf{k} + \textbf{q}})^2$), one can surmise that $H^{\rm Cooper}_{\rm pair} + H^{\rm screened}_{\rm Coulomb}$ is a constant such that $\langle H^{\rm Cooper}_{\rm pair} + H^{\rm screened}_{\rm Coulomb}\rangle = \langle V_{\textbf{k},\textbf{k}'}\rangle = V' < 0$ where for convenience, $|V'| = V > 0$ is defined to cause the attraction between electrons. Based on this approximation, BCS moved on to construct an elegant Hamiltonian~\cite{bard},
\begin {eqnarray}
H^{\rm BCS}_{\rm reduced} = 2\sum_{k > k_{\rm F}}\epsilon_{\textbf{k}}b^*_{\textbf{k}}b_{\textbf{k}} + 2\sum_{k < k_{\rm F}}|\epsilon_{\textbf{k}}|b_{\textbf{k}}b^*_{\textbf{k}} - V\sum_{\textbf{k},\textbf{k}'}b^*_{\textbf{k}'}b_{\textbf{k}}, \label{eq:6} 
\end {eqnarray}  
by assuming that each Cooper pair is a boson-like particle, and these (ground state) pairs only form in the vicinity of Fermi energy ($E_{\rm F}$). But Cooper pairs are composed of electrons, and these pairs are not bosons in a real physical sense due to Cooper-pair formation mechanism explained earlier between Eq.~(\ref{eq:3}) and Eq.~(\ref{eq:4}). This means that, a Cooper pair as an independent boson-like entity cannot obey Fermi-Dirac (FDS) or Bose-Einstein (BES) statistics~\cite{bard}. However, the electrons in Cooper pairs do obey FDS, which will also be addressed later.  

In Eq.~(\ref{eq:6}), $b^*_{\textbf{k}}$ and $b_{\textbf{k}}$ creates and annihilates a Cooper pair (two electrons), respectively, hence the factor 2, $b_{\textbf{k}} = c_{-\textbf{k}\downarrow}c_{\textbf{k}\uparrow}$, $b^*_{\textbf{k}} = c^*_{\textbf{k}\uparrow}c^*_{-\textbf{k}\downarrow}$, $[b_{\textbf{k}},b^*_{\textbf{k}'}] = (1 - n_{\textbf{k}\uparrow} - n_{-\textbf{k}\downarrow})\delta_{\textbf{k}\textbf{k}'}$, $[b_{\textbf{k}},b_{\textbf{k}'}] = 0$, $\{b_{\textbf{k}},b_{\textbf{k}'}\} = 2b_{\textbf{k}}b_{\textbf{k}'}(1-\delta_{\textbf{k}\textbf{k}'})$, $n_{\textbf{k}\sigma} = c^*_{\textbf{k}\sigma}c_{\textbf{k}\sigma}$, $c^*_{\textbf{k}}c_{\textbf{k}'} = 1 - c_{\textbf{k}'}c^*_{\textbf{k}}$, $c_{\textbf{k}}c_{\textbf{k}'} = - c_{\textbf{k}'}c_{\textbf{k}}$, $n_{\textbf{k}}c_{\textbf{k}'} = c_{\textbf{k}'}n_{\textbf{k}}$ and $n_{\textbf{k}}c_{\textbf{k}} = 0$. Some of these identities have been used to obtain Eq.~(\ref{eq:6}) from Eqs.~(\ref{eq:1}) and~(\ref{eq:3}).

Next, BCS used some guessed wavefunctions to finally derive the renowned gap equation that determines the superconductor transition temperature~\cite{bard},
\begin {eqnarray}
\Delta_{\rm BCS} = k_{\rm B}T^{\rm BCS}_{\rm sc} = 1.14\hbar\omega\exp{\bigg[-\frac{1}{N(0)V}\bigg]}, \label{eq:7} 
\end {eqnarray}  
where $N(0)$ is the density of states at Fermi level, normalized by letting $E_{\rm F} = 0$. Of course, Eq.~(\ref{eq:7}) strictly satisfies the condition, $k_{\rm B}T^{\rm BCS}_{\rm sc} \ll \hbar\omega$, which allows us to treat $\langle V_{\textbf{k},\textbf{k}'}\rangle = V$ as a constant earlier. Therefore, by definition, BCS-theory obeying systems are nothing but Onnes-type weakly-coupled conventional superconductors. Importantly, the isotope effect is captured by the term $\omega^2 = k^{\rm interaction}_{\rm const}/M_{\rm ion}$ where $k^{\rm interaction}_{\rm const}$ is the ion-ion interaction potential constant, while $M_{\rm ion}$ denotes ion mass. The isotope effect has been shown to be canceled in Eq.~(\ref{eq:3}), and therefore $V$ is immune to any changes in $\omega$~\cite{bardpines}. Even though high $T^{\rm BCS}_{\rm sc}$ superconductors can be predicted from these parameters, $\omega$, $N(0)$ and $V$, but one is left groping for the microscopic physics needed to understand the changes in $V$ for materials with different atoms and compositions. This inadequacy (including the ones highlighted earlier) never imply that BCS Hamiltonian is doomed for unconventional superconductors.
 
As a consequence, our primary aim here is to formally show why Cooper pairs can be strongly bounded (with high superconductor gap), regardless of their coherence lengths by reconstructing $V^{\rm attraction}_{\textbf{k},\textbf{k}'}$ as a function of doping parameter where $V^{\rm attraction}_{\textbf{k},\textbf{k}'}$ is not a constant, and the attraction is uniquely between two electrons. We then go on to derive the FDS for the electrons that have formed Cooper pairs between $T = 0$ and $T_{\rm sc}$ such that these pairs are still boson-like. Finally, we prove the existence of (a) doping($x$)-dependent $T_{\rm sc}$, controlled by Cooper-pair (or superfluid) density ($n^{\rm Cooper}_{\rm pair}$), and (b) QPT$^{\rm >0K}$ giving rise to the $C_{\rm v}$ discontinuity at the critical point ($T_{\rm sc}$). The above objectives shall be properly covered in the following section.      

\section*{2. Theoretical results}

Our strategy here is to first prove the existence of a generalized potential operator such that e:ph interaction can either induce the e:e attraction via phonon exchange to give rise to a superconducting phase, or can play its role in the form of electron-ion (e:ion) attraction to activate the usual e:e Coulomb repulsion. This repulsion can either produce the strange metallic phase (if the energy levels are still degenerate) or an insulator (due to Mott or band gap). Subsequently, we derive the FDS for Cooper pairs to understand their excitation probability below $T_{\rm sc}$ by counting the Cooper-pair electrons, instead of Cooper pairs. Finally, we invoke the above generalized potential operator to show that this potential allows QPT$^{\rm >0K}$ to exist, which is responsible for the phase transition between superconductivity and normal state property.

\subsubsection*{2.1. Strongly bounded Cooper pairs}

Earlier, stronger e:ph interaction is shown to have the physical capability to produce two types of e:e interactions (see Eqs.~(\ref{eq:4}) and~(\ref{eq:5})), one is the expected Coulomb repulsion, while the other is due to Cooper attraction. The said repulsion and attraction between electrons refer to $(\hbar\omega_{\textbf{q}})^2 - (E_{\textbf{k}} - E_{\textbf{k} + \textbf{q}})^2 > 0$ and $-(\hbar\omega_{\textbf{q}})^2 + (E_{\textbf{k}} - E_{\textbf{k} + \textbf{q}})^2 < 0$, respectively. Hence, $V_{\textbf{k},\textbf{k}'}$ should accommodate both Eqs.~(\ref{eq:4}) and~(\ref{eq:5}), which means,
\begin {eqnarray}
V_{\textbf{k},\textbf{k}'} = \bigg\{^{V^{\rm repulsion}_{\textbf{k},\textbf{k}'}~ :~ {\rm for}~ \big[(\hbar\omega_{\textbf{q}})^2 - (E_{\textbf{k}} - E_{\textbf{k} + \textbf{q}})^2 ~>~ 0\big]}_{V^{\rm attraction}_{\textbf{k},\textbf{k}'}~ :~ {\rm for}~\big[-(\hbar\omega_{\textbf{q}})^2 + (E_{\textbf{k}} - E_{\textbf{k} + \textbf{q}})^2 ~<~ 0\big]}, \label{eq:8} 
\end {eqnarray}
where stronger repulsion and attraction can be achieved by a larger magnitude of $(E_{\textbf{k}} - E_{\textbf{k} + \textbf{q}})^2$. For example, if a particular electron (that is not part of a Cooper pair) absorbs or emits $(E_{\textbf{k}} - E_{\textbf{k} \pm \textbf{q}})$ a relatively high energy phonon, $\hbar\omega_{\textbf{q}'}$, then $(\hbar\omega_{\textbf{q}})^2 - (E_{\textbf{k}} - E_{\textbf{k} \pm \textbf{q}'})^2 < (\hbar\omega_{\textbf{q}})^2 - (E_{\textbf{k}} - E_{\textbf{k} \pm \textbf{q}})^2$ that readily leads to large $V^{\rm repulsion}_{\textbf{k},\textbf{k}'}$ where $\hbar\omega_{\textbf{q}'} > \hbar\omega_{\textbf{q}}$. 

Similarly, for a Cooper electron to be strongly bounded to another Cooper electron, one also requires $(E_{\textbf{k}} - E_{\textbf{k} + \textbf{q}})^2$ to be large. However, an additional requirement is needed such that $E_{\textbf{k}} - E_{\textbf{k} + \textbf{q}}$ for electron 1 should also lead to $E_{\textbf{k}'} - E_{\textbf{k}' - \textbf{q}}$ for electron 2 where electron 1 and 2 form a Cooper pair. The spins can be suppressed because it is straightforward to note that $\textbf{k}$ and $\textbf{k}'$ refer to $\sigma$ and $\sigma'$, respectively, and high-energy phonon exchange between electron 1 and 2 gives rise to high binding energy for Cooper pairs, and consequently a larger $V^{\rm attraction}_{\textbf{k},\textbf{k}'}$. Even if electron 1 and 2 has been considered as a single entity (boson-like) by BCS, but in our formalism, these electrons (1 and 2) are treated as individuals, as they should be. Treating each Cooper-pair electrons as an individual particle does not violate the original boson-like Cooper pair formation mechanism (bounded due to $V^{\rm attraction}_{\textbf{k},\textbf{k}'}$), which can be shown to be valid from the following BCS identity,      
\begin {eqnarray}
[b_{\textbf{k}},b^*_{\textbf{k}'}] = (- n_{\textbf{k}\uparrow}\delta_{\uparrow\uparrow} - n_{-\textbf{k}\downarrow}\delta_{\downarrow\downarrow} + c^*_{\textbf{k}'\uparrow}c_{-\textbf{k}\downarrow}\delta_{\uparrow\downarrow} + c^*_{-\textbf{k}'\downarrow}c_{\textbf{k}\uparrow}\delta_{\uparrow\downarrow})\delta_{\textbf{k}\textbf{k}'}, \label{eq:9} 
\end {eqnarray}
where $n_{\textbf{k}\uparrow}$ and $n_{-\textbf{k}\downarrow}$ are the unpaired electron numbers, and therefore
\begin {eqnarray}
[b_{\textbf{k}},b^*_{\textbf{k}'}] = (1 - n_{\textbf{k}\uparrow} - n_{-\textbf{k}\downarrow})\delta_{\textbf{k}\textbf{k}'}, \label{eq:10} 
\end {eqnarray}
after letting $c^*_{\textbf{k}'\uparrow}c_{-\textbf{k}\downarrow} = 0$ and $c^*_{-\textbf{k}'\downarrow}c_{\textbf{k}\uparrow} = 0$ due to $\delta_{\uparrow\downarrow} = 0$ where $c^*$ and $c$ are the respective creation and annihilation operators for individual (unpaired) electrons. Here, the number of Cooper pairs has to be $1 - n_{\textbf{k}\uparrow} - n_{-\textbf{k}\downarrow}$ for $\textbf{k} = \textbf{k}'$ as given in Eq.~(\ref{eq:10}). Apparently, the creation and annihilation of Cooper pairs ($b^*$ and $b$) require the creation and annihilation of individual electrons such that they can be paired.

Since our Cooper pairs have large binding energies compared to conventional superconductors, one has no other option but to supersede the BCS approximation, $(\hbar\omega_{\textbf{q}})^2 \gg (E_{\textbf{k}} - E_{\textbf{k} + \textbf{q}})^2$ with $(\hbar\omega_{\textbf{q}})^2 > (E_{\textbf{k}} - E_{\textbf{k} + \textbf{q}})^2$, which implies $V^{\rm attraction}_{\textbf{k},\textbf{k}'}$ can no longer be considered a constant (denoted earlier by $V$). Our next aim is to properly define $V^{\rm attraction}_{\textbf{k},\textbf{k}'}$ in such a way that it is also a function of atomic energy-level spacing, $\xi$. We first make use of the energy-level spacing renormalization group method~\cite{aop} to renormalize $V_{\textbf{k},\textbf{k}'}$ and $|g_{\textbf{k},\textbf{k}'}|^2$ given in Eq.~(\ref{eq:4}) to obtain (also after the sign-switch transformation), 
\begin {eqnarray}
&&V^{\rm attraction}_{\textbf{k},\textbf{k}'} = |g_{\textbf{k},\textbf{k}'}|^2\frac{2\hbar\omega_{\textbf{q}}e^{\frac{1}{2}\lambda(\xi - E^0_{\rm F})}}{-\big[\hbar\omega_{\textbf{q}}e^{\frac{1}{2}\lambda(\xi - E^0_{\rm F})}\big]^2 + \big[E_{\textbf{k}} - E_{\textbf{k} + \textbf{q}}\big]^2}, \label{eq:11} \\&& |g_{\textbf{k},\textbf{k}'}|^2 = \frac{1}{\Omega}\frac{e^2}{\epsilon\big[|\textbf{q}|^2+ K^2_se^{\lambda(-\xi + E^0_{\rm F})}\big]}\frac{1}{2}\hbar\omega_{\textbf{q}}e^{\frac{1}{2}\lambda(\xi - E^0_{\rm F})}, \label{eq:12} 
\end {eqnarray}
where $\Omega$ is the volume in momentum space, and our renormalization method employed here can be exactly mapped onto the Shankar's wavenumber-dependent renormalization technique~\cite{shank}. Now we invoke the large binding-energy condition that validates $(\hbar\omega_{\textbf{q}})^2 > (E_{\textbf{k}} - E_{\textbf{k} + \textbf{q}})^2$ such that $\Delta(\hbar\omega_{\textbf{q}})^2 \ll \Delta(E_{\textbf{k}} - E_{\textbf{k} + \textbf{q}})^2$ where $\Delta$ denotes the change in the stated variables due to temperature and doping applicable for cuprate superconductors. This inequality ($\ll$) is understandable as the change in $T_{\rm sc}$ ($\Delta T_{\rm sc}$) due to isotope effect ($\Delta\hbar\omega_{\textbf{q}}$) is minute, compared to $\Delta T_{\rm sc}$ as a result of $\Delta(E_{\textbf{k}} - E_{\textbf{k} + \textbf{q}})$ where $(E_{\textbf{k}} - E_{\textbf{k} + \textbf{q}}) \cong k_{\rm B}T_{\rm sc}$.

Physically, $\hbar\omega_{\textbf{q}}e^{\frac{1}{2}\lambda(\xi - E^0_{\rm F})} \leq \big[E_{\textbf{k}} - E_{\textbf{k} + \textbf{q}}\big]$ is never allowed for two reasons---(i) $\leq$ $\rightarrow$ $=$ requires perfect phonon exchange, and on the other hand, (ii) $\leq$ $\rightarrow$ $<$ violates the second law of thermodynamics. Equation~(\ref{eq:11}) tells us that one can increase $T_{\rm sc}$ by increasing $\big[E_{\textbf{k}} - E_{\textbf{k} + \textbf{q}}\big]$ to be close to $\hbar\omega_{\textbf{q}}e^{\frac{1}{2}\lambda(\xi - E^0_{\rm F})}$, which in turn implies large $V^{\rm attraction}_{\textbf{k},\textbf{k}'}$. To obtain large $\big[E_{\textbf{k}} - E_{\textbf{k} + \textbf{q}}\big]$ however, one also needs large $\big[E_{\textbf{k}'} - E_{\textbf{k}' - \textbf{q}}\big]$ so as to set the stage for high Cooper-pair-formation probability. Obviously, this probability can be enhanced if the density of states at the Fermi level is large, which is nothing but what is required from the BCS gap equation, Eq.~(\ref{eq:7}) through $N(0)$.

One of the most important implications of BCS Cooper-pair formation with respect to Eq.~(\ref{eq:11}) (with or without extension) is the existence of a proper supercurrent-flow mechanism~\cite{bard} such that the scattering rates ($1/\tau$), induced by the electron-phonon ($1/\tau_{\rm e:ph}$), electron-electron ($1/\tau_{\rm e:e}$) and spin-disorder ($1/\tau_{\rm sd}$) scattering processes are literally zero. As explained earlier, $E_{\textbf{k},\sigma} - E_{\textbf{k} + \textbf{q},\sigma}$ for the first electron require the second electron to satisfy $E_{\textbf{k}',\sigma'} - E_{\textbf{k}' - \textbf{q},\sigma'}$, which imply $1/\tau_{\rm e:ph} = 0 = 1/\tau_{\rm e:e}$ because the Cooper-pair formation mechanism obviously requires these systematic changes in momenta, $\textbf{k} \rightarrow \textbf{k} + \textbf{q}$ and $\textbf{k}' \rightarrow \textbf{k}' - \textbf{q}$, and also due to $|V^{\rm attraction}_{\textbf{k},\textbf{k}'}| > V^{\rm repulsion}_{\textbf{k},\textbf{k}'}$. Here, we do not need $V^{\rm repulsion}_{\textbf{k},\textbf{k}'} = 0$ to obtain $1/\tau_{\rm e:e} = 0 = 1/\tau_{\rm e:ph}$ because the e:e and e:ph scattering processes are entirely responsible for the above stated momentum-change in the first and second electrons. 

Now, $1/\tau_{\rm sd}$ is also zero because the first electron with spin, $\sigma$ do not scatter the second electron with spin, $\sigma'$ if these two electrons momenta change in this way, $[\textbf{k},\sigma \rightarrow \textbf{k} + \textbf{q},\sigma]_{\rm electron1}$ and $[\textbf{k}',\sigma' \rightarrow \textbf{k}' - \textbf{q},\sigma']_{\rm electron2}$, respectively. In other words, the above scattering processes are not zero, but they are entirely responsible to initiate the required phonon-mediated momentum-change for Cooper-pair formation, and therefore, these scattering processes do not cause resistance, which means, $1/\tau_{\rm e:e} = 0 = 1/\tau_{\rm e:ph} = 1/\tau_{\rm sd}$. The above exposition also applies to triplet pairing symmetry where we just need to allow an additional criterion, $\sigma = \sigma'$ for such pairing.  

On the contrary, the supercurrent-flow mechanism on the basis of (i) spin-fluctuation pairing as campaigned by Scalapino~\cite{scala} and (ii) Bose-condensed resonating valence bonds as advocated by Anderson and Baskaran~\cite{ander,bask} remain hidden because their proposals assume that supercurrent is an `automatic' consequence once the superglue~\cite{scala} or Bose-Einstein condensation~\cite{ander} is identified. Hence, their theories ignore this essential requirement, $1/\tau_{\rm e:ph} = 0 = 1/\tau_{\rm e:e} = 1/\tau_{\rm sd}$ for superconductivity. Moreover, if the spins are required to fluctuate in order to induce electron-pair formation~\cite{scala}, then one is also left puzzled as to the reason why and how $1/\tau_{\rm sd}$ can be zero. Details on $1/\tau_{\rm sd}$ within the transport theory of ferromagnets can be found in Ref.~\cite{pssb}. Therefore, alternative theories should first settle the primary issues of superconductivity with respect to (a) supercurrent-flow mechanism, (b) Meissner effect, (c) specific heat capacity jump at the critical temperature and (d) doping-dependent critical temperature and resistivity for $T > T_{\rm sc}$ before nominating any possible theories for contention.
  
Apart from that, magnetic interaction can and should exist in one form or another in cuprates or other unconventional superconductors because cuprates are antiferromagnets (with different types of atoms arranged in different sequences in $ab$-plane compared to $c$-axis), and moreover, Cooper-pair binding energy also depends on the electron's spin (see Eqs.~(\ref{eq:3}),~(\ref{eq:9}) and~(\ref{eq:10})). But, as we have said many times now, and as unambiguously shown above, the magnetic interaction is unlikely to be the cause for electron-pairing.
  
In summary, even though BCS Hamiltonian is not perfect, but its correctness and consistency on the basis of well established microscopic physics are unparalleled (compared to other alternatives), even when the Cooper-pair binding energy is made to be large by replacing the BCS condition $(\hbar\omega_{\textbf{q}})^2 \gg (E_{\textbf{k}} - E_{\textbf{k} + \textbf{q}})^2$ with $(\hbar\omega_{\textbf{q}})^2 > (E_{\textbf{k}} - E_{\textbf{k} + \textbf{q}})^2$. Hence, one should be convinced by now that there is no such thing as BCS Hamiltonian cannot handle high $T_{\rm sc}$ materials, or other unconventional superconductors. Rightly so, we have stopped identifying $T_{\rm sc}$ as $T^{\rm BCS}_{\rm sc}$ or $T^{\rm BM}_{\rm sc}$.

\subsubsection*{2.2. Fermi-Dirac statistics for Cooper-pair electrons}

In the preceding sub-section, we have proven that the BCS attraction operator is a special case (see Eq.~(\ref{eq:8})) that allows the formation of Cooper pairs, composed of two individual electrons, coupled attractively by means of emission and absorption of phonons. This picture of looking at each Cooper pair electrons individually shall allow us to derive the FDS for both Cooper-pair and unpaired electrons. In particular, each Cooper pair is not considered as a single boson-like entity (because it is neither a boson nor a two-fermion entity in a real physical sense), but as two individual electrons with changes in energies, $E_{\textbf{k}\uparrow} - E_{\textbf{k}'\uparrow}$ and $E_{-\textbf{k}\downarrow} - E_{-\textbf{k}'\downarrow}$ such that they are coupled attractively due to phonon exchange. Additionally, Cooper-pair formation satisfies the conservation of crystal momentum where $\textbf{k} + (-\textbf{k}) = \textbf{k}' + (-\textbf{k}') = \textbf{k} + \textbf{q} + (-\textbf{k} - \textbf{q}) = 0$. 

As a consequence of the above picture, one can readily exploit the FDS to understand the excitation probability of Cooper pairs. Here, the excitation of Cooper-pair electrons simply means breaking up of Cooper pairs. The restrictive conditions for both unpaired and Cooper-pair electrons are given by,
\begin {eqnarray}
&&\sum^{\infty}_{i = 1}N_i = N, \label{eq:13} \\&& \sum^{\infty}_{i = 1}N_i\bigg[E + \frac{1}{2}|\langle V^{\rm attraction}_{\textbf{k},\textbf{k}'}\rangle|\bigg]_i = E + |\langle V^{\rm attraction}_{\textbf{k},\textbf{k}'}\rangle|, \label{eq:14}  
\end {eqnarray}
where $N$ denotes the total number of electrons, including Cooper-pair electrons, and the factor 1/2 in Eq.~(\ref{eq:14}) avoids counting $|\langle V^{\rm attraction}_{\textbf{k},\textbf{k}'}\rangle|$ twice because it takes two to activate the attraction. Moreover, $E > |\langle V^{\rm attraction}_{\textbf{k},\textbf{k}'}\rangle|$, or at least $E > 0$ is always true physically because not all electrons can form Cooper pairs---it would be outrageous if we were to assume that localized core electrons can and will form Cooper pairs. If $T \rightarrow T_{\rm sc}$, then $|\langle V^{\rm attraction}_{\textbf{k},\textbf{k}'}\rangle| \rightarrow 0$ and at the same time $E \rightarrow E^{\rm nsc}_{\rm phase}$ where $E^{\rm nsc}_{\rm phase}$ is the total energy in the non-superconducting phase (above $T_{\rm sc}$). 

The BCS dispersion relation, $E_{\textbf{k}'\uparrow}$ $=$ $\sqrt{\epsilon^2_{\textbf{k}'} + \epsilon^2_{0}}$ differs from Eq.~(\ref{eq:14}) because the said dispersion is only valid for energies near the Fermi level~\cite{bard} such that $E_{\textbf{k}'\uparrow}$ $+$ $E_{\textbf{k}''\uparrow}$ $=$ $2\epsilon_0 = 2\Delta^{\rm BCS}$ because $\epsilon_{\textbf{k}'} \rightarrow 0$ and $\epsilon_{\textbf{k}''} \rightarrow 0$ where $E_{\textbf{k}''\uparrow}$ $=$ $\sqrt{\epsilon^2_{\textbf{k}''} + \epsilon^2_{0}}$. The limit $\epsilon_{\textbf{k}} \rightarrow 0$ implies all electrons near Fermi level are paired. In addition, if $\textbf{k}'_{\uparrow}$ and $-\textbf{k}''_{\downarrow}$ are unoccupied, then $\textbf{k}''_{\uparrow}$ and $-\textbf{k}'_{\downarrow}$ are occupied. In other words, the BCS dispersion relation only considers electrons that will form Cooper pairs near Fermi level. Anyway, after following the standard procedure~\cite{grif} using Eqs.~(\ref{eq:13}) and~(\ref{eq:14}), one should be able to derive the sought-after statistics for both Cooper-pair- and unpaired-electrons,
\begin {eqnarray}
f^{\rm Cooper}_{\rm FDS} &=& \frac{1}{e^{(E + |\langle V^{\rm attraction}_{\textbf{k},\textbf{k}'}\rangle| - E^{0}_{\rm F})/k_{\rm B}T} + 1}, \\&\cong& \exp{\bigg[\frac{E^{0}_{\rm F} - |\langle V^{\rm attraction}_{\textbf{k},\textbf{k}'}\rangle| - E}{k_{\rm B}T}\bigg]}. \label{eq:15}  
\end {eqnarray}
As anticipated, the above statistics correctly guides us to this fact---the probability for breaking up Cooper-pair electrons ($f^{\rm Cooper}_{\rm FDS}$) is diminished (as it should be) for decreasing $T$ or increasing $|\langle V^{\rm attraction}_{\textbf{k},\textbf{k}'}\rangle|$. Above $T_{\rm sc}$, $|\langle V^{\rm attraction}_{\textbf{k},\textbf{k}'}\rangle| \rightarrow 0$, $E \rightarrow E^{\rm nsc}_{\rm phase}$ and therefore $f^{\rm Cooper}_{\rm FDS} \rightarrow f^{\rm standard}_{\rm FDS}$. Here, $f^{\rm standard}_{\rm FDS}$ can also be written as a function of $E^{\rm nsc}_{\rm phase} = E' + |\langle V^{\rm repulsion}_{\textbf{k},\textbf{k}'}\rangle|$ where $E'$ denotes the non-interacting energy, which correctly points to the fact that large e:e repulsion increases the energy-level spacing, leading to smaller excitation probability, $f^{\rm standard}_{\rm FDS}$ for the usual (unpaired) electrons. Hence, the above transformation, $f^{\rm Cooper}_{\rm FDS} \rightarrow f^{\rm standard}_{\rm FDS}$ is physically valid such that $f^{\rm Cooper}_{\rm FDS}$ is not only unique for Cooper-pair electrons, but it is also not \textit{ad hoc}.  

\subsubsection*{2.3. Ionization energy approximation}

The origin of $T_{\rm sc}$ has been exposed earlier (beyond Eq.~(\ref{eq:7})) on the basis of $V_{\textbf{k},\textbf{k}'}$ defined in Eq.~(\ref{eq:8}) such that Eq.~(\ref{eq:3}) is a special case. The phase transition from $V_{\textbf{k},\textbf{k}'}^{\rm repulsion}$ to $V_{\textbf{k},\textbf{k}'}^{\rm attraction}$ or vice versa is activated by QPT$^{\rm >0K}$. We also have extended the BCS Hamiltonian to capture cuprate superconductors by allowing the Cooper pairs to be strongly bounded (by requiring $(\hbar\omega_{\textbf{q}})^2 > (E_{\textbf{k}} - E_{\textbf{k} + \textbf{q}})^2$). 

In the subsequent sub-sections, we shall make the microscopic origin for the existence of QPT$^{\rm >0K}$ due to $V_{\textbf{k},\textbf{k}'}^{\rm repulsion} \rightarrow V_{\textbf{k},\textbf{k}'}^{\rm attraction}$ transformation explicit. This result should help us to understand the doping-dependent $T_{\rm sc}$ effect. This notorious effect shall be tackled by exploiting the only theory that allows us to do so consistently without resorting to any variationally adjustable parameters and guessed wavefunctions. The theory is known as the ionization energy theory (IET) that relies on the ionization energy approximation~\cite{pra,aop}. The downside of our tactic on the basis of IET is that it cannot be used to predict $T_{\rm sc}$ for a given material because the analysis is at best abstract at the lowest (or operator) level, which is already apparent from our analysis presented earlier. It so happens that, regardless of the approach employed, one cannot predict $T_{\rm sc}$ microscopically, which shall be exposed on the fly later. 

However, the advantage of using IET is profound as our analysis are not only microscopically precise, unambiguous and consistent, but can also be used to discover (i) exactly where does $V_{\textbf{k},\textbf{k}'}^{\rm attraction}$ originate from (see Eq.~(\ref{eq:8})), (ii) that the Cooper-pair formation is indeed the correct mechanism for superconductivity in cuprates (see the analysis between Eqs.~(\ref{eq:4}) and~(\ref{eq:5})), and (iii) the possibility to pin down the precise microscopic mechanism for doping-dependent $T_{\rm sc}$, which is controlled by $n^{\rm Cooper}_{\rm pair}$. Apart from these points ((i), (ii) and (iii)), one should also note this---only the lowest-level analysis can unequivocally prove whether a theory is free of any internal inconsistency. 

We now briefly introduce IET. More details on IET and its formalism in different context have been reported in Refs.~\cite{pra,aop,jsnm,jsnm2,a1}. The term ionization energy ($\xi$) in IET is precisely the energy-level spacing ($\xi$) of a given system where $\xi_{\rm system}$ varies systematically if one carries out a systematic substitutional doping. Here, the ionization energy approximation can be employed to determine $\xi_{\rm system}$ from its constituent atoms where the approximation reads~\cite{jsnm2},
\begin {eqnarray}
\xi_{\rm system} = \sum_j\sum^z_i\frac{1}{z}\xi_{i,j}(\texttt{X}^{i+}_j), \label{eq:16a}
\end {eqnarray}
where each $\texttt{X}_j$ represents a particular atom in a given system such that $j > 1$ implies that there are more than one type of atoms, while $i$ counts the valence electrons for each type of constituent atoms. Here, $\xi_{i,j}(\texttt{X}^{i+}_j)$ denotes the energy level spacing for constituent atom, $\texttt{X}^{i+}_j$ and its energy-level spacing or ionization energy values can be directly obtained from any validated databases listed in Refs.~\cite{ral,win}.    

If a system is made up of free electrons (Fermi gas) or weakly-interacting fermions (Fermi liquid), then IET or its approximation is literally useless because $\xi$ is either zero or it is a nonzero constant ($\xi = \xi_{\rm irr} \neq 0$). In semiconductors and insulators however, $\xi$ is neither zero nor a constant, and it is known as the trivially relevant energy-level spacing (or $\xi_{\rm triv} \neq 0$), and it can refer to a band or Mott-Hubbard or molecular gap. In the early days of our investigation~\cite{jsnm2}, we have discovered that the strange metallic phase in cuprates above $T_{\rm sc}$ has an anomalous gap, which turned out to be a nontrivial energy-level spacing, $\xi^{\rm non}_{\rm triv} \neq 0$~\cite{jsnm}. It is nontrivial because the energy levels are degenerate (gapless) and yet, $\xi^{\rm non}_{\rm triv} \neq 0$, and therefore, $\xi^{\rm non}_{\rm triv}$ determines the electron transition probability between different orthogonal and degenerate wavefunctions~\cite{jsnm}. Our immediate aim now is to formally get $\xi^{\rm non}_{\rm triv}$ on board or into Eq.~(\ref{eq:5}), and then analyze QPT$^{\rm >0K}$ with respect to doping (or changing chemical composition).

\subsubsection*{2.4. Finite-temperature quantum phase transition}

One can invoke $\xi^{\rm non}_{\rm triv}$ by realizing that these normalized wavefunctions, $\psi_{\textbf{k},\sigma}$, $\psi_{\textbf{k}',\sigma'}$, $\psi_{\textbf{k}+\textbf{q},\sigma}$ and $\psi_{\textbf{k}'-\textbf{q},\sigma'}$ are orthogonal to each other by definition and they can be degenerate. The existence of strange metallic phase in cuprates above $T_{\rm sc}$ necessitates one to write $|E_{\textbf{k}} - E_{\textbf{k} + \textbf{q}}| = \xi^{\rm non}_{\rm triv} = \xi$ and $|E_{\textbf{k}'} - E_{\textbf{k}' - \textbf{q}'}| = \xi'$ following Ref.~\cite{jsnm} where $\xi = \xi'$ if $\textbf{q} = \textbf{q}'$. From here onwards, our notation for $\xi^{\rm non}_{\rm triv}$ reverts to $\xi$ for convenience because we focus only on cuprates. Introducing this substitution into Eqs.~(\ref{eq:4}) and~(\ref{eq:5}) gives us the large effective Cooper attraction between electrons,
\begin {eqnarray}
\langle V^{\rm attraction}_{\textbf{k},\textbf{k}'}\rangle + \langle V^{\rm repulsion}_{\textbf{k},\textbf{k}'}\rangle &=& \bigg\langle\frac{2\hbar\omega_{\textbf{q}}|g_{\textbf{k},\textbf{k}'}|^2e^{\frac{1}{2}\lambda(\xi - E^0_{\rm F})}}{-\big[\hbar\omega_{\textbf{q}}e^{\frac{1}{2}\lambda(\xi - E^0_{\rm F})}\big]^2 + \xi^2} + \frac{2\hbar\omega_{\textbf{q}}|g_{\textbf{k},\textbf{k}'}|^2e^{\frac{1}{2}\lambda(\xi - E^0_{\rm F})}}{\big[\hbar\omega_{\textbf{q}}e^{\frac{1}{2}\lambda(\xi - E^0_{\rm F})}\big]^2 - \xi^2}\bigg\rangle, \label{eq:16} \\&=& \langle\psi_{\textbf{k}'-\textbf{q},\sigma'}\psi_{\textbf{k}',\sigma'}|H^{\rm Cooper}_{\rm pair} + H^{\rm screened}_{\rm Coulomb}|\psi_{\textbf{k},\sigma}\psi_{\textbf{k}+\textbf{q},\sigma}\rangle, \label{eq:17} 
\end {eqnarray}
where $H^{\rm screened}_{\rm Coulomb} = V^{\rm repulsion}_{\textbf{k},\textbf{k}'}$, which has been proven in Ref.~\cite{aop} in the absence of very large effective mass effect. Recall that this substitution is not applicable for conventional superconductors because the normal state of BCS superconductors satisfy Fermi gas ($\xi = 0$) or Fermi liquid ($\xi \rightarrow  \xi_{\rm irr} \neq 0$). In other words, the electron transition probability between the wavefunctions defined in Eq.~(\ref{eq:16}) for BCS superconductors is always one because there is no energy barrier to cross over. In any case, we have gotten what we needed, namely, Eq.~(\ref{eq:16}), which is in a suitable form to extract the necessary information on QPT$^{\rm >0K}$. 

It is straightforward to deduce the following features from Eq.~(\ref{eq:16}). In the strange metallic phase, $\langle V^{\rm attraction}_{\textbf{k},\textbf{k}'}\rangle = 0$ and $\langle V^{\rm attraction}_{\textbf{k},\textbf{k}'}\rangle$ is maximum, and for $T_{\rm sc}$, QPT$^{\rm >0K}$ is activated such that QPT$^{\rm >0K}$ = QPT$^{T_{\rm sc}}$ when $\langle V^{\rm attraction}_{\textbf{k},\textbf{k}'}\rangle \neq 0$ and $|\langle V^{\rm attraction}_{\textbf{k},\textbf{k}'}\rangle| > \langle V^{\rm repulsion}_{\textbf{k},\textbf{k}'}\rangle$. Interestingly, we may have preformed Cooper pairs for $T > T_{\rm sc}$ if and only if $\langle V^{\rm attraction}_{\textbf{k},\textbf{k}'}\rangle \neq 0$ and $|\langle V^{\rm attraction}_{\textbf{k},\textbf{k}'}\rangle| < \langle V^{\rm repulsion}_{\textbf{k},\textbf{k}'}\rangle$ where $n^{\rm Cooper}_{\rm pair}$ is not sufficiently high to activate QPT$^{T_{\rm sc}}$. In addition, the existence of preformed Cooper pairs does not imply that QPT$^{T_{\rm sc}}$ or the transition to superconducting phase is inevitable. For $T_{\rm sc} > T \rightarrow 0$, $\langle V^{\rm repulsion}_{\textbf{k},\textbf{k}'}\rangle$ should approach zero, while $|\langle V^{\rm attraction}_{\textbf{k},\textbf{k}'}\rangle|$ $\rightarrow$ maximum where $\langle V^{\rm repulsion}_{\textbf{k},\textbf{k}'}\rangle$ can be zero because both $\langle V^{\rm repulsion}_{\textbf{k},\textbf{k}'}\rangle$ and $\langle V^{\rm attraction}_{\textbf{k},\textbf{k}'}\rangle$ refer only to electrons (near Fermi level) that can form Cooper pairs. In contrast, the restrictive condition given in Eq.~(\ref{eq:16}) counts all the electrons in a given system, and therefore, $E > 0$ is mandatory because of core electrons contribution.

We now discuss the consequence of changing $\xi$ (due to doping) on $\langle V^{\rm repulsion}_{\textbf{k},\textbf{k}'}\rangle$ and $\langle V^{\rm attraction}_{\textbf{k},\textbf{k}'}\rangle$. The effect of varying $\xi$ that comes from an exponential term in the numerator is more or less canceled by the same term in the denominator (see Eq.~(\ref{eq:16})). Whereas, $|g_{\textbf{k},\textbf{k}'}|$ (defined in Eq.~(\ref{eq:12})) is proportional to $\xi$, which is as it should be because increasing $\xi$ should results in the amplification of $\langle V^{\rm repulsion}_{\textbf{k},\textbf{k}'}\rangle$ (see Eq.~(\ref{eq:16})). This repulsion is also further enhanced if $\xi^2$ in the denominator becomes large provided that $\big[\hbar\omega_{\textbf{q}}e^{\frac{1}{2}\lambda(\xi - E^0_{\rm F})}\big]^2$ increases slower than $\xi^2$ or $\Delta\big[\hbar\omega_{\textbf{q}}e^{\frac{1}{2}\lambda(\xi - E^0_{\rm F})}\big]^2$ $<$ $\Delta\xi^2$. Another essential point here is that the repulsion never requires a second electron to emit ($\textbf{k} - \textbf{q}$) the absorbed phonon ($\textbf{k} + \textbf{q}$) by the first electron (see Eq.~(\ref{eq:3})), and this leads to a maximum repulsion. 

In contrast, the relationship between $\xi$ (due to doping) and $\langle V^{\rm attraction}_{\textbf{k},\textbf{k}'}\rangle$ is not so straightforward because $\xi$ does not determine the formation of Cooper pairs, or $\xi$ is not the cause for superconductivity. As explained earlier, superconductivity occurs if the second electron happens to emit the phonon absorbed by the first electron. If this scenario (systematic emission and absorption of phonons) occurs exclusively for large number of electrons, then (and only then), $\langle V^{\rm attraction}_{\textbf{k},\textbf{k}'}\rangle$ becomes relevant in such a way that these Cooper pairs shall lead to QPT$^{T_{\rm sc}}$, provided that $|\langle V^{\rm attraction}_{\textbf{k},\textbf{k}'}\rangle| > \langle V^{\rm repulsion}_{\textbf{k},\textbf{k}'}\rangle$. 

This means that, predicting a superconducting material is a hard problem because the source for the existence of $\langle V^{\rm attraction}_{\textbf{k},\textbf{k}'}\rangle$ is not predictable by definition, even within the BCS Hamiltonian. For example, we do not know exactly what observable (or measurable) parameter causes (or induces) this scenario---when one electron absorbs a phonon, the second electron happens to emit an identical phonon (note this point). What we know (from BCS Hamiltonian) is that we need large $N(0)$ to increase $T_{\rm sc}$, but it is not responsible for superconductivity. This is similar to $\xi$, which is also not responsible for superconductivity. This means that, if a particular system is a superconductor, then we can invoke $\xi$ to evaluate the changes to $\langle V^{\rm attraction}_{\textbf{k},\textbf{k}'}\rangle$ or $T_{\rm sc}$ with respect to doping. 

Similar to the relationship between $\langle V^{\rm repulsion}_{\textbf{k},\textbf{k}'}\rangle$ and $\xi$, one can also amplify $\langle V^{\rm attraction}_{\textbf{k},\textbf{k}'}\rangle$ by increasing $\xi$, again provided that $\Delta\big[\hbar\omega_{\textbf{q}}e^{\frac{1}{2}\lambda(\xi - E^0_{\rm F})}\big]^2$ $<$ $\Delta\xi^2$. Consequently, both $\langle V^{\rm repulsion}_{\textbf{k},\textbf{k}'}\rangle$ and $\langle V^{\rm attraction}_{\textbf{k},\textbf{k}'}\rangle$ are proportional to $\xi$ but due to Cooper-pair formation, $\langle V^{\rm attraction}_{\textbf{k},\textbf{k}'}\rangle$ becomes the dominant interaction below $T_{\rm sc}$, and therefore, one can immediately predict an interesting outcome here---increasing $\xi$ does not indefinitely increases $\langle V^{\rm attraction}_{\textbf{k},\textbf{k}'}\rangle$. In other words, there exists an optimal doping or $\xi_{\rm optimum}$ with maximum $\langle V^{\rm attraction}_{\textbf{k},\textbf{k}'}\rangle$ or $T_{\rm sc}$, controlled by $n^{\rm Cooper}_{\rm pair}$. As we have pointed out earlier, it is not possible to determine why $n^{\rm Cooper}_{\rm pair}$ below $T_{\rm sc}$ changes with doping, namely, why $n^{\rm Cooper}_{\rm pair}$ is maximum for $\xi_{\rm optimum}$, and otherwise for $\xi < \xi_{\rm optimum}$ or $\xi > \xi_{\rm optimum}$. For example, large $\xi$ is required to obtain high $T_{\rm sc}$ (due to strongly bounded Cooper pairs), but large $\xi$ is not responsible for the formation of Cooper pairs, for the same reason large $N(0)$ is required to increase $T_{\rm sc}$, but large $N(0)$ is not the cause for superconductivity. But never mind, at least, we now know exactly what one cannot know within BCS Hamiltonian, with or without extension.

In summary, we have extended the BCS Hamiltonian to be applicable to cuprates without internal inconsistency (by means of operator-level analysis), explained why and how $T_{\rm sc}$ can be high (due to Cooper-pair formation in the presence of large $\xi$), and have shown the existence of quantum phase transition (denoted by QPT$^{T_{\rm sc}}$) and optimal doping concentration ($x_{\rm optimum}$). Here, QPT$^{T_{\rm sc}}$ and $x_{\rm optimum}$ exist due to temperature- and doping-dependent $n^{\rm Cooper}_{\rm pair}$, and also because of the competition between $\langle V^{\rm repulsion}_{\textbf{k},\textbf{k}'}\rangle$ and $\langle V^{\rm attraction}_{\textbf{k},\textbf{k}'}\rangle$  (see Eq.~(\ref{eq:16})). Finally, we have established that predicting a superconductor material is next to impossible because we do not even know what parameter induces the formation of Cooper pairs. 

\subsubsection*{2.5. Specific heat capacity at critical point}

Earlier, we have shown that a Cooper-pair binding energy can be large if the pair is formed in the presence of large $\xi$, provided the energy levels are still degenerate and $|\langle V^{\rm attraction}_{\textbf{k},\textbf{k}'}\rangle| > \langle V^{\rm repulsion}_{\textbf{k},\textbf{k}'}\rangle$. However, the role played by $\xi$ when $|\langle V^{\rm attraction}_{\textbf{k},\textbf{k}'}\rangle| > \langle V^{\rm repulsion}_{\textbf{k},\textbf{k}'}\rangle$ remains ambiguous or hidden (see Eq.~(\ref{eq:16})), and therefore, experimentally not verifiable because unlike $n^{\rm Cooper}_{\rm pair}$, $\xi$ is not responsible for superconductivity, at least not directly. This means that, we need to find a way to show that $\xi$ is indeed responsible for strongly bounded Cooper pairs. To achieve this, we make use of the finite-temperature quantum phase transition theory developed in Ref.~\cite{ijp} to address QPT$^{T_{\rm sc}}$ in cuprates. In particular, we require $\xi$ to be unequivocal in determining the reason why $C_{\rm v}$ is discontinuous at $T_{\rm sc}$, as well as why $C_{\rm v}$ jumps up (not down) such that $C^{\rm paired}_{\rm v}(T_{\rm sc}) > C^{\rm unpaired}_{\rm v}(T_{\rm sc})$ where `paired' and `unpaired' refer to Cooper pairs and unpaired electrons, respectively.

We stress that the validity and correctness of $\xi$ above $T_{\rm sc}$ in cuprates are irrefutable~\cite{jsnm,jsnm2}, and we will not reproduce them here. Briefly though, $\xi$ has been proven to give unambiguous microscopic explanations on the doping-dependent electrodynamics above $T_{\rm sc}$. Here in this last sub-section before we wrap up, we shall attempt to provide a direct evidence that $\xi$ also plays a leading role in the phase transition from the normal to superconductor state by associating $\xi$ to the notion of QPT$^{T_{\rm sc}}$. This evidence implies that $\xi$ is indeed responsible for the formation of strongly bounded Cooper pairs. Recall here that $\xi$ is either zero or denotes a irrelevant constant in conventional superconductors where `irrelevant' means $\xi$ does not play any role on the electrodynamics of a given solid~\cite{aop}.

Above $T_{\rm sc}$, the specific heat capacity is given by $C^{\rm unpaired}_{\rm v}(T > T_{\rm sc})$, while $C^{\rm paired}_{\rm v}(T < T_{\rm sc})$ is obviously valid below $T_{\rm sc}$. At the critical point (for $T = T_{\rm sc}$) however, both $C^{\rm unpaired}_{\rm v}(T > T_{\rm sc})$ and $C^{\rm paired}_{\rm v}(T < T_{\rm sc})$ are invalid as already proven in Ref.~\cite{ijp} by studying the solidification and melting processes. In particular, when $T = T_{\rm sc}$, Cooper-pair formation gives rise to QPT$^{T_{\rm sc}}$ such that $\langle V^{\rm repulsion}_{\textbf{k},\textbf{k}'}\rangle$ is no longer a constant (because $\langle V^{\rm repulsion}_{\textbf{k},\textbf{k}'}\rangle > |\langle V^{\rm attraction}_{\textbf{k},\textbf{k}'}\rangle| \rightarrow \langle V^{\rm repulsion}_{\textbf{k},\textbf{k}'}\rangle < |\langle V^{\rm attraction}_{\textbf{k},\textbf{k}'}\rangle|$), and this transformation is only possible if we allow $\xi$ to change significantly at $T_{\rm sc}$. 

Note this carefully, unlike solidification or melting process~\cite{ijp}, the change in $\xi$ at $T_{\rm sc}$ in cuprates does not (in any way) refers to changing $|E_{\textbf{k}} - E_{\textbf{k} + \textbf{q}}| = \xi$ or $|E_{\textbf{k}'} - E_{\textbf{k}' - \textbf{q}}| = \xi$, but refers to increasing or decreasing $i$ in $\sum_i|E_{\textbf{k}} - E_{\textbf{k} + \textbf{q}}|_i = \sum_i\xi_i$ or $\sum_i|E_{\textbf{k}'} - E_{\textbf{k}' - \textbf{q}}|_i = \sum_i\xi_i$. The summation here counts the number of unpaired electrons that controls the magnitude of $\langle V^{\rm repulsion}_{\textbf{k},\textbf{k}'}\rangle$. Actually, we have the option to choose either to count the unpaired or the paired electrons (Cooper pairs), but we prefer to count the unpaired electrons because we know exactly how to relate $\xi$ to $C^{\rm unpaired}_{\rm v}(T > T_{\rm sc})$ as already proven in Refs.~\cite{aop,ijp}. Thus, we have no other choice but to avoid counting the paired electrons because we do not know the relation between $\xi$ and $C^{\rm paired}_{\rm v}(T < T_{\rm sc})$ that determines $|\langle V^{\rm attraction}_{\textbf{k},\textbf{k}'}\rangle|$. Warning: if the change in $\xi$ is due to doping, then one requires the energy-level spacing itself, namely, $|E_{\textbf{k}} - E_{\textbf{k} + \textbf{q}}| = \xi$ to change as discussed earlier and in Ref.~\cite{ijp}.
\begin{figure}
\begin{center}
\scalebox{0.35}{\includegraphics{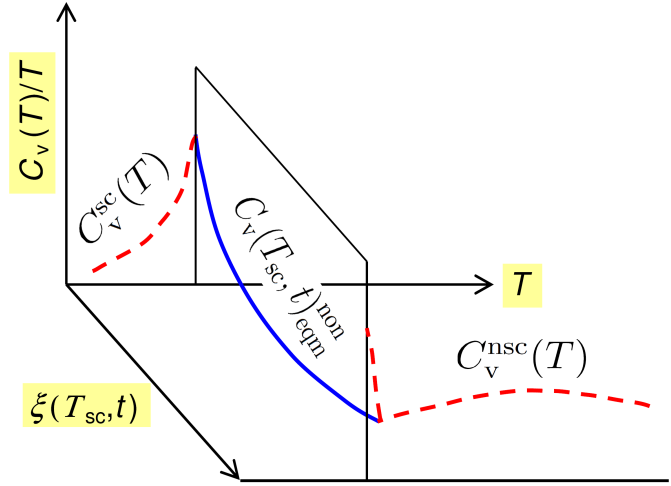}}
\caption{Temperature($T$)- and time($t$)-dependent specific heat capacity for $T > T_{\rm sc}$, $T = T_{\rm sc}$ and $T < T_{\rm sc}$ denoted respectively by $C^{\rm nsc}_{\rm v}(T)$, $C_{\rm v}(T_{\rm sc},t)^{\rm non}_{\rm eqm}$ and $C^{\rm sc}_{\rm v}(T)$. The experimental specific heat capacity ($C_{\rm v}(T)$), which is observable, is sketched by the dashed line where the microscopic process (Cooper-pair formation) that occurs at the critical point ($T_{\rm sc}$) stayed hidden. The solid line reveals the effect of Cooper-pair formation at a constant temperature, $T = T_{\rm sc}$ captured by the $t$-dependent $\xi(T_{\rm sc},t)$ (see Eqs.~(\ref{eq:18}) and~(\ref{eq:19})). The dashed lines (above and below $T_{\rm sc}$) are defined by $C^{\rm nsc}_{\rm v}(T)$ and $C^{\rm sc}_{\rm v}(T)$, respectively, where the discontinuity at $T_{\rm sc}$ gives rise to a divergent specific heat based on classical thermodynamics. The solid line ($C_{\rm v}(T_{\rm sc},t)^{\rm non}_{\rm eqm}$) is obtained from the quantum thermodynamical approach formulated in Ref.~\cite{ijp}.}
\label{fig:1}
\end{center}
\end{figure}

Having explained that, we can now exploit Eq.~(28) derived in Ref.~\cite{ijp} such that the specific heat change right at the critical point during the transformation from non-superconducting (nsc or unpaired) to superconducting (sc or paired) state is given by the non-equilibrium specific heat capacity,
\begin {eqnarray}
C_{\rm v}(T_{\rm sc},t)^{\rm non}_{\rm eqm} = C_{\rm v}(T_{\rm sc})\exp{\bigg[-\frac{3}{2}\lambda\sum_iJ^{\rm nsc}_i\xi_{\rm nsc}\bigg]}, \label{eq:18} 
\end {eqnarray}
where $C_{\rm v}(T_{\rm sc},t)^{\rm non}_{\rm eqm}$ is also the dynamic (or time($t$)-dependent) specific heat at a constant temperature, $T_{\rm sc}$, $\sum_iJ^{\rm nsc}_i$ sums the decreasing unpaired electron density that measures the `strength' of $\xi_{\rm nsc}$, which is a constant for each electron. The strength of $\xi_{\rm nsc}$ is maximum if all electrons are unpaired ($\sum_iJ^{\rm nsc}_i = 1$). Furthermore, we have $C_{\rm v}(T_{\rm sc})$ that can either transform into $C^{\rm nsc}_{\rm v}(T)$ if $T$ is increased from $T < T_{\rm sc}$ to $T > T_{\rm sc}$, or $C_{\rm v}(T_{\rm sc}) \rightarrow C^{\rm sc}_{\rm v}(T)$ if one reduces the temperature from $T > T_{\rm sc}$ to $T < T_{\rm sc}$. For the first transformation (sc to nsc), one has,  
\begin {eqnarray}
C_{\rm v}(T_{\rm sc},t)^{\rm non}_{\rm eqm} = C_{\rm v}(T_{\rm sc})\exp{\bigg[-\frac{3}{2}\lambda\sum_i(1 - J^{\rm sc}_i)\xi_{\rm nsc}\bigg]}, \label{eq:19} 
\end {eqnarray}
where
\begin {eqnarray}
\sum_iJ^{\rm nsc}_i = \frac{\sum_i\big[n^{\rm unpaired}_{\rm electron}\big]_i}{N^{\rm unpaired}_{\rm total}}, ~~~ \sum_iJ^{\rm sc}_i = \frac{\sum_i\big[n^{\rm paired}_{\rm electron}\big]_i}{N^{\rm paired}_{\rm total}}, \label{eq:20} 
\end {eqnarray}
and $\sum_i(1 - J^{\rm sc}_i)$ counts the decreasing number of paired electrons such that $\sum_i(1 - J^{\rm sc}_i) = 1$ if all electrons are unpaired. For paired electrons, we still count them individually. It is now straightforward to observe (see Fig.~\ref{fig:1}) that when nsc transforms into sc, $\sum_iJ^{\rm nsc}_i \rightarrow 0$ or $\sum_iJ^{\rm sc}_i \rightarrow 1$ and $C_{\rm v}(T_{\rm sc}) \rightarrow C^{\rm sc}_{\rm v}(T)$ that give us the relation, $C^{\rm sc}_{\rm v}(T) > C^{\rm nsc}_{\rm v}(T)$ because of decreasing magnitude of $\sum_iJ^{\rm nsc}_i\xi_{\rm nsc}$ (see Eqs.~(\ref{eq:18}) and~(\ref{eq:20})). Alternatively, from Eqs.~(\ref{eq:19}) and~(\ref{eq:20}), when sc transforms into nsc, we have $\sum_i(1 - J^{\rm nsc}_i) \rightarrow 1$ and $C_{\rm v}(T_{\rm sc}) \rightarrow C^{\rm nsc}_{\rm v}(T)$ (due to increasing $\sum_iJ^{\rm nsc}_i\xi_{\rm nsc}$), which eventually lead us to $C^{\rm nsc}_{\rm v}(T) < C^{\rm sc}_{\rm v}(T)$.

Hence, solely on the basis of BCS Hamiltonian extended within IET formalism, we have proven that $C_{\rm v}(T)$ jumps up at $T_{\rm sc}$ if we approach $T_{\rm sc}$ from $T > T_{\rm sc}$ to $T < T_{\rm sc}$. On the other hand, $C_{\rm v}(T)$ jumps down at $T_{\rm sc}$ if one reaches the normal state from the superconductor phase by increasing the temperature. These jumps refer to the same measured discontinuity in $C_{\rm v}(T)/T$ at $T_{\rm sc}$, which has been sketched in Fig.~\ref{fig:1} following Ref.~\cite{loram} for proper visualization. Note this, the specific heat capacity depicted in Fig.~\ref{fig:1} is found to jump upward at $T_{\rm sc}$, determined entirely from the first principles, which is in agreement with experimental data reported by Loram \textit{et al.}~\cite{loram}.

In contrast, for conventional superconductors, BCS theory in its original form has been used to correctly derive the specific-heat jump for $T = T_{\rm sc}$ by calculating the difference in free energies (see Eq.~(36.9) on page 306 in Ref.~\cite{gor}). In particular, the said jump can be understood from this relation, $C^{\rm sc}_{\rm v}(T_{\rm sc}) = C^{\rm nsc}_{\rm v}(T_{\rm sc}) + \alpha T_{\rm sc}$ where $C^{\rm sc}_{\rm v}(T_{\rm sc}) > C^{\rm nsc}_{\rm v}(T_{\rm sc})$, and $\alpha$ here symbolically represents a collection of constants~\cite{gor}. However, the transition from $C^{\rm sc}_{\rm v}(T_{\rm sc})$ to $C^{\rm nsc}_{\rm v}(T_{\rm sc})$ or vice versa that defines the existence of $C_{\rm v}(T_{\rm sc},t)^{\rm non}_{\rm eqm}$ remains unknown within the original BCS theory because we do not know the explicit function for $\Delta_{\rm BCS}(T_{\rm sc},t)$, thus, the jump is always assumed to be discontinuous on the basis of Landau phase transition theory~\cite{gor}. 

In order to derive $\Delta_{\rm BCS}(T_{\rm sc},t)$, we need to find the parameter that controls the Cooper-pair formation. For example, we need to discover the parameter that is responsible for this phenomenon---when one electron absorbs a phonon, the second electron happens to emit an identical phonon that has been pointed out earlier. Within IET formalism, we can ignore $\Delta_{\rm BCS}(T_{\rm sc},t)$, and instead, one can construct the function, $C_{\rm v}(T_{\rm sc},t)^{\rm non}_{\rm eqm}$ (see Eqs.~(\ref{eq:18}),~(\ref{eq:19}) and~(\ref{eq:20})) to expose that the transition to superconducting phase is a continuous process, which is as it should be from quantum thermodynamics point of view~\cite{ijp}. This means that, our lack of knowledge on the function, $\Delta_{\rm BCS}(T_{\rm sc},t)$ does not imply that it does not exist, for the same reason we cannot assume $C_{\rm v}(T_{\rm sc},t)^{\rm non}_{\rm eqm}$ does not exist just because $C_{\rm v}(T)$ measurements do not indicate any continuous process at $T_{\rm sc}$~\cite{ptp}. Unfortunately, IET is by definition not applicable for conventional superconductors because their normal states satisfy Fermi-liquid metallic properties, and therefore, $\Delta_{\rm BCS}(T_{\rm sc},t)$ cannot be approximated from IET.

\section*{3. Conclusions}

Even though BCS Hamiltonian is an idealized model of superconductivity, but it has been properly extended and generalized here to capture the physics of cuprate superconductors without any internal inconsistency. Anisotropic resistivity, $d$-wave pairing, doping-dependent superconductor gap or $T_{\rm sc}$ and the pseudogap above $T_{\rm sc}$ do not rule out the formation of Cooper-pairs in its original form. On the other hand, alternative theories such as spin-fluctuation induced pairing~\cite{pines,scala}, resonating valence bond theory~\cite{ander,bask}, and bipolarons~\cite{schra,alex,chak} do not reliably handle superconductivity in general for these two solid reasons---the microscopic mechanism for zero electric and magnetic fields (\textbf{E} = 0 = \textbf{B}) supercurrent flow below $T_{\rm sc}$ (1) demands the phonons to Bose condense in one form or another, or (2) neglects the phonon-induced scattering (if (1) is not required), without any microscopically relevant physical justifications. 

For example, the phonon-induced scattering mechanism between phonons and paired (or Bose-condensed) polarons, spin-induced electron pairs, spinons and holons were not properly eliminated in these alternative supercurrent-flowing mechanisms (see the first paragraph in introduction). On the contrary, Cooper pairs and its conduction mechanism below $T_{\rm sc}$ and for \textbf{E} = 0 = \textbf{B} are well-defined in such a way that phonons play an active role by directly mediating the formation of Cooper pairs, and therefore, the e:ph scattering effect is readily and properly eliminated. In addition, these phonons are not required to Bose-condense, even below $T_{\rm sc}$. Rightly so, we have decided to extend the well thought-out BCS Hamiltonian, in its original form, to cuprate superconductors with a much stronger Cooper pairs regardless of their coherence lengths. Also note this, sufficient operator-level analysis at the critical point is completely missing in these alternative theories.

We have developed a comprehensive microscopic physics of superconductivity for cuprates based on the BCS Hamiltonian, which also allowed us to invoke QPT$^{\rm >0K}$ to explain the existence of $T_{\rm sc}$ and the possibilities of phonon readjustment and preformed Cooper pairs above $T_{\rm sc}$. Most importantly, our approach is entirely based on first principles such that we did not resort to any guessed functions or fitting-parameter tactics to justify the validity and correctness of BCS Hamiltonian in cuprates. Hence, it is not an exaggeration to declare that BCS Hamiltonian could be applicable to all types of superconductors due to its unambiguous superiority in taming phonons, and in handling Cooper-pair formation and supercurrent flow microscopically and consistently at the operator level in accordance with the most relevant and important experiments. 

The relevant experiments are associated to (1) isotope effect, (2) dissipationless supercurrent flow below $T_{\rm sc}$, including the Meissner effect, (3) Specific heat below $T_{\rm sc}$ ($C^{\rm sc}_{\rm v}(T)$), (4) doping- and temperature-dependent superconductor gap (on the basis of Eqs.~(\ref{eq:15}) and~(\ref{eq:16})), (5) doping-dependent normal state resistivity above $T_{\rm sc}$ (see Ref.~\cite{jsnm2}), (6) the existence of strange metallic phase above $T_{\rm sc}$ (see Ref.~\cite{jsnm}), and (7) the discontinuous specific heat capacity at $T_{\rm sc}$ (see Eqs.~(\ref{eq:18}),~(\ref{eq:19}) and~(\ref{eq:20})). This specific heat discontinuity due to QPT$^{T_{\rm sc}}$ opens up the path to stretch the energy-level spacing influence in the strange metallic phase to the superconducting state. Bardeen, Cooper and Schrieffer have already addressed points (1) to (3) decades ago, while the extended BCS Hamiltonian within IET has been used to tackle points (4) and (7).

\section*{Acknowledgments}

Even though Madam Sebastiammal Savarimuthu and Mr Arulsamy Innasimuthu were disappointed with my repeated failures to tackle superconductivity since 1999, but they are still fascinated by the `floating-magnet' experiment to the extent that they did not hesitate to reinstate their support.

\end{document}